\documentstyle[prd,aps,amssymb,preprint,epsfig]{revtex}
\begin{document}


\def\Bbar{{\bar B}}
\def\Dbar{{\bar D}}
\def\calB{{\cal B}}
\def\calO{{\cal O}}
\def\Lambdabar{{\bar\Lambda}}


\def\etal{{\it et al.}}
\def\ibid#1#2#3{{\it ibid.} {\bf #1}, #3 (#2)}
\def\epjc#1#2#3{Eur. Phys. J. C {\bf #1}, #3 (#2)}
\def\ijmpa#1#2#3{Int. J. Mod. Phys. A {\bf #1}, #3 (#2)}
\def\jhep#1#2#3{J. High Energy Phys. {\bf #1}, #3 (#2)}
\def\mpl#1#2#3{Mod. Phys. Lett. A {\bf #1}, #3 (#2)}
\def\npb#1#2#3{Nucl. Phys. {\bf B#1}, #3 (#2)}
\def\plb#1#2#3{Phys. Lett. B {\bf #1}, #3 (#2)}
\def\prd#1#2#3{Phys. Rev. D {\bf #1}, #3 (#2)}
\def\prl#1#2#3{Phys. Rev. Lett. {\bf #1}, #3 (#2)}
\def\rep#1#2#3{Phys. Rep. {\bf #1}, #3 (#2)}
\def\zpc#1#2#3{Z. Phys. {\bf #1}, #3 (#2)}
\def\ibid#1#2#3{{\it ibid}. {\bf #1}, #3 (#2)}


\title{Nonleptonic $B$ decays into a charmed tensor meson}
\author{Jong-Phil Lee\footnote{e-mail: jplee@phya.yonsei.ac.kr}}
\address{Department of Physics and IPAP, Yonsei University, Seoul, 120-749, Korea}

\tighten
\maketitle

\begin{abstract}

In the framework of the factorization and the heavy quark effective theory, 
$B\to D_2^*\pi$ modes are analyzed.
We adopt the result from the QCD sum rule calculation for the hadronic matrix
elements at leading order of $\Lambda_{\rm QCD}/m_Q$ and $\alpha_s$.
The QCD sum rule results are well compatible with the current data, with the
prediction on the branching ratios 
$\calB(\Bbar^0\to D_2^{*+}\pi^-)=8.94\times 10^{-4}$ and
$\calB(B^-\to D_2^{*0}\pi^-)=9.53\times 10^{-4}$ for $N_C^{\rm eff}=2$.
We give constraints on the interception $\tau(1)$ and the slope parameter 
$\rho^2$ of the leading Isgur-Wise function from the experimental bounds.
It is argued that the observation of nonzero $\calB(\Bbar^0\to D_2^{*0}\pi^0)$
directly measures the nonfactorizable effects.

\end{abstract}
\pagebreak

\section{Introduction}

The advent of the $B$-factory era in KEK and SLAC opens a new chance for very
suppressed $B$ decays.
Nonleptonic two-body decays into tensor $(T)$ mesons, among them, deserve much
attention nowadays.
Experimental data on them provides only upper bounds of the branching ratios.
Two-body hadronic $B$ decays involving a tensor meson $T$ in the final state
has long been studied \cite{Katoch,Munoz,Oh} using the non-relativistic
quark model of Isgur, Scora, Grinstein, and Wise (ISGW) \cite{ISGW} with the
factorization ansatz.
Their predictions for the branching ratios are rather small while the
preliminary results from the Belle Collaboration indicate that the branching
ratios for $B\to PT$ ($P=$ pseudoscalar) may not be very small compared
to $B\to PP$ modes \cite{Belle}.
Recently, both charmed and charmless $B\to P(V)T$ ($V=$ vector) decays were 
updated \cite{jplee} using ISGW2 model \cite{ISGW2} which is an HQET-based 
improvement of its original model.
They are in many respects complementary to $B\to P(V)$ decays.
In a theoretical point of view, more reliable description of $B\to T$ transition
is required.
\par
The biggest obstacle in theoretical predictions is the hadronic matrix elements.
The factorization hypothesis is a widely accepted assumption.
Recently, it is pointed out that the factorization parameter $a_2$ is process
dependent and there is a nonzero strong phase difference between color-allowed
and color-suppressed decay modes, based on the first observation of 
$\Bbar^0\to D^{(*)0}\pi^0$ by Belle and CLEO \cite{Abe,Xing}.
It is thus very interesting to see what happens in $B\to D_2^{*}\pi$.
The vector meson $D_1$ and the charmed tensor $D_2^*$ are members of a doublet 
$(1^+,2^+)$ with
$j_l^P=3^+/2$, while $(D'_0,D'_1)$ corresponds to $(0^+,1^+)$ with 
$j_l^P=1^+/2$.
The neutral decay mode $\Bbar^0\to D_2^{*0}\pi^0$ is dominated by the 
color-suppressed internal $W$-emission diagram.
Within the factorization (we will not consider the small contributions from 
the $W$- exchange diagram for simplicity),  
the decay amplitude is proportional to $\langle 0|V-A|T\rangle$.
It can be easily shown, however, that such a factorized term vanishes \cite{Oh}.
This is a good advantage of tensor mesons in the final state because the 
decay amplitude is greatly simplified.
Given the factorization assumption, therefore, the hadronic uncertainties are
condensed to the $B\to T$ transition matrix elements.
On the other hand, the observation of sizable branching ratio for 
$\Bbar^0\to D_2^{*0}\pi^0$ would provide the direct information on the
nonfactorizable effects \cite{Oh,Neubert,Diehl}. 
This is another benefit of studying the production of tensor mesons.
\par
Present bounds from the experiments are \cite{PDG}
\begin{eqnarray}
\calB(B^+\to \Dbar_2^{*0}\pi^+)<1.3\times 10^{-3}~,\nonumber\\
\calB(B^0\to \Dbar_2^{*-}\pi^+)<2.2\times 10^{-3}~.
\end{eqnarray}
\par
In this paper, we analyze the two-body $B$ decays into a charmed tensor $D_2^*$.
Heavy quark effective theory (HQET) is an appropriate framework in this process.
In HQET, $B\to D_2^*$ transition matrix element is parametrized by one
universal Isgur-Wise (IW) function at leading order of $\Lambda_{\rm QCD}/m_Q$,
where $m_Q$ is the heavy quark mass.
An extensive study of the leading and subleading IW function in semileptonic
decays is found in \cite{Leibovich}.
In general, the IW function depends on the velocity transfer $y\equiv v\cdot v'$
where $v(v')$ is the 4-velocity of $B(D)$.
Typically, kinematically allowed range of $y-1$ is very small in $B\to D$
transition.
It is customary to parametrize the IW function $\tau(y)$ in terms of its
interception $\tau(1)$ and the slope parameter $\rho^2$, and expand in $(y-1)$.
The branching ratio is directly proportional to $|\tau(1)|^2$.
Unlike the groundstate to groundstate transition, the heavy quark symmetry (HQS)
does not guarantee the normalization of $\tau(1)$ in $B\to D_2^*$.
This is because at zero recoil ground- to excited-state transition is 
suppressed by $\calO(\Lambda_{\rm QCD}/m_Q)$, due to the HQS.
\par
Though the HQET is very economic and allows a systematic expansion of 
$\Lambda_{\rm QCD}/m_Q$, one still needs some nonperturbative methods to
evaluate the IW function.
At present work, we adopt the QCD sum rule results for the $B\to D_2^*$ 
leading IW function \cite{Huang}.
QCD sum rule is among the most reliable nonperturbative methods \cite{sumrule}.
It takes into account the nontrivial QCD vacuum which is parametrized by various
vacuum condensates in order to describe the nonperturbative nature.  
In QCD sum rule, hadronic observables can be calculated by evaluating two- or 
three-point correlation functions.  
The hadronic currents for constructing the correlation 
functions are expressed by the interpolating fields.  
In describing the excited $D$ mesons of $(1^+,2^+)$ states, the transverse
covariant derivative is included in the interpolating fields \cite{Huang}.
\par
In the next section, the decay amplitudes are given within the factorization 
and the QCD sum rule results for the leading IW function are summarized.
Section III contains the numerical results and discussions.
The QCD sum rule results are compared with the ISGW2 model predictions.
Possible next-to-leading order corrections are also discussed.
We summarize in Sec.\ IV.

\section{Hadronic matrix elements and QCD sum rules}

The effective weak Hamiltonian for $B\to D_2^*\pi$ is
\begin{equation}
{\cal H}_{\rm eff}=\frac{G_F}{\sqrt{2}}V_{cb}V_{ud}^*\Big[
 c_1(\mu)({\bar d}u)({\bar c}b)+c_2(\mu)({\bar c}u)({\bar d}b)+\cdots\Big]~,
\end{equation}
where $({\bar q}_1q_2)={\bar q}_1\gamma^\mu(1-\gamma_5)q_2$ and $c_i(\mu)$ are
the Wilson coefficients.
\par
Within the factorization framework, the decay rate amplitudes are given by
\begin{mathletters}
\label{amp}
\begin{eqnarray}
{\cal A}_{+-}&\equiv&
  {\cal A}(\Bbar^0\to D_2^{*+}\pi^-)={\cal T}+{\cal E}~,\\
{\cal A}_{00}&\equiv&
  {\cal A}(\Bbar^0\to D_2^{*0}\pi^0)=\frac{1}{\sqrt{2}}(-{\cal C}+{\cal E})~,\\
{\cal A}_{0-}&\equiv&
  {\cal A}(B^-\to D_2^{*0}\pi^-)={\cal T}+{\cal C}~,
\end{eqnarray}
\end{mathletters}
where 
\begin{mathletters}\label{topology}
\begin{eqnarray}
{\cal T}&=&\frac{G_F}{\sqrt{2}}V_{cb}V^*_{ud}
 \langle\pi^-|(\bar{d}u)_{V-A}|0\rangle
 \langle D_2^{*+}|(\bar{c}b)_{V-A}|\Bbar^0\rangle a_1 ~,\\
{\cal C}&=&\frac{G_F}{\sqrt{2}}V_{cb}V^*_{ud}
 \langle D_2^{*0}|(\bar{c}u)_{V-A}|0\rangle
 \langle\pi^0|(\bar{d}b)_{V-A}|\Bbar^0\rangle a_2~,\\
{\cal E}&=&\frac{G_F}{\sqrt{2}}V_{cb}V^*_{ud}
 \langle\Bbar^0|(\bar{d}b)_{V-A}|0\rangle
 \langle D_2^{*0}\pi^0|(\bar{c}u)_{V-A}|0\rangle a_2~,
\end{eqnarray}
\end{mathletters}
are the color-allowed external $W$-emission, color-suppressed internal $W$-
emission and $W$-exchange amplitudes, respectively.
The coefficients $a_{1,2}$ are defined by
\begin{equation}
a_1\equiv c_1+\xi c_2~,~~~a_2\equiv c_2+\xi c_1~,
\end{equation}
where $\xi\equiv 1/N_C^{\rm eff}$ and $N_C^{\rm eff}$ is the effective number
of color.
Note that (\ref{amp}) satisfies the isospin triangle relation
\begin{equation}
{\cal A}_{+-}=\sqrt{2}{\cal A}_{00}+{\cal A}_{0-}~.
\label{triangle}
\end{equation}
The factorized matrix elements are parametrized as
\begin{mathletters}
\begin{eqnarray}
\langle 0 | A^{\mu} | P \rangle &=& i f_P p_P^{\mu} ~,  \\
\langle T | j^{\mu} | B \rangle &=& i h(m_{P}^2) \epsilon^{\mu \nu
\rho \sigma} \epsilon^*_{\nu \alpha} p_B^{\alpha} (p_B
+p_T)_{\rho} (p_B -p_T)_{\sigma} + k(m_{P}^2) \epsilon^{* \mu \nu}
(p_B)_{\nu}  \nonumber \\
&\mbox{}&  + \epsilon^*_{\alpha \beta} p_B^{\alpha} p_B^{\beta} [
b_+(m_{P}^2) (p_B +p_T)^{\mu} +b_-(m_{P}^2) (p_B -p_T)^{\mu} ]~,
\label{formfactor}
\end{eqnarray}
\end{mathletters}
where $j^{\mu} = V^{\mu} -A^{\mu}$ and  $V^{\mu}$ ($A^{\mu}$)
denote a vector (an axial-vector) current.
Here $f_P$
denotes the decay constant of the relevant pseudoscalar meson, and
$h(m_{P}^2)$, $k(m_{P}^2)$, $b_+(m_{P}^2)$, $b_-(m_{P}^2)$ 
express the form factors for the $B \rightarrow T$ transition.
\par
We will neglect the internal $W$-exchange diagram ${\cal E}$ for simplicity.
And the color-suppressed internal $W$-emission diagrma ${\cal C}$ is 
forbbiden, because \cite{Oh}
\begin{equation}
{\cal C}\sim \langle 0|j^\mu|T(p,\lambda)\rangle
\sim p_\nu\epsilon^{\mu\nu}(p,\lambda)+p^\mu\epsilon^\nu_\nu(p,\lambda)=0~.
\end{equation}
Thus, the decay mode $\Bbar^0\to D_2^{*0}\pi^0$ is not allowed in the 
factorization scheme.
\par
In the heavy quark limit where $m_Q\to\infty$, all the form factors are
expressed by one universal Isgur-Wise (IW) function $\tau(y\equiv v\cdot v')$
\cite{Huang}:
\begin{mathletters}
\begin{eqnarray}
h(y)&=&\frac{\tau(y)}{2m_B\sqrt{m_B m_T}}~,\\
k(y)&=&\sqrt{\frac{m_T}{m_B}}(1+y)\tau(y)~,\\
b_+(y)&=&-\frac{\tau(y)}{2m_B\sqrt{m_B m_T}}~,\\
b_-(y)&=&\frac{\tau(y)}{2m_B\sqrt{m_B m_T}}~,
\end{eqnarray}
\end{mathletters}
where $v$ ($v'$) denotes the four velocity of $B$ ($T$).
\par
At this point, nonperturbative methods are needed to evaluate $\tau(y)$.
We use the QCD sum rule results at leading order of $\Lambda_{\rm QCD}/m_Q$.
The QCD sum rule result for $\tau(y)$ is \cite{Huang}
\begin{eqnarray}\label{tau}
\tau(y)f_{-,1/2}f_{+,3/2}~e^{-(\Lambdabar_{-,1/2}+\Lambdabar_{+,3/2})/M}
&=&
\frac{1}{2\pi^2(y+1)^3}\int_0^{\omega_c} d\omega_+\omega_+^3 e^{-\omega_+/M}
\nonumber\\
&&-\frac{1}{12}m_0^2\frac{\langle {\bar q}q\rangle}{M}
  -\frac{1}{3\times 2^5\pi}\langle\alpha_s GG\rangle\frac{y+5}{(y+1)^2}~,
\end{eqnarray}
where
\begin{mathletters}\label{f}
\begin{eqnarray}
f_{-,1/2}^2~e^{-2\Lambdabar_{-,1/2}/M}&=&
\frac{3}{16\pi^2}\int_0^{\omega_{c0}}\omega^2e^{-\omega/M}d\omega
 -\frac{1}{2}\langle{\bar q}q\rangle\Bigg(1-\frac{m_0^2}{4M^2}\Bigg)~,\\
f_{+,3/2}^2~e^{-2\Lambdabar_{+,3/2}/M}&=&
\frac{1}{2^6\pi^2}\int_0^{\omega_{c2}}\omega^4e^{-\omega/M}d\omega
 -\frac{1}{12}m_0^2\langle{\bar q}q\rangle-\frac{1}{2^5\pi}
  \langle\alpha_sGG\rangle M~,
\end{eqnarray}
\end{mathletters}
from which $f_{\pm,3/2(1/2)}$ and $\Lambdabar_{\pm,3/2(1/2)}$ are determined.
Here $\langle{\bar q}q\rangle$ and $\langle\alpha_s GG\rangle$ are vacuum
condensates, $m_0^2=0.8~{\rm GeV}^2$, and $M$ is the Borel parameter.
The continuum thresholds $\omega_c$, $\omega_{c0}$, and $\omega_{c2}$ are
adjustable parameters for the numerical analysis.
Typically, the kinematically allowed range of $y$ is very narrow,
$1\le y\lesssim 1.3$.
It is therefore convenient to approximate $\tau(y)$ linearly as
\begin{equation}
\tau(y)=\tau(1)[1-\rho^2(y-1)]~.
\label{linear}
\end{equation}
The QCD sum rule predicts from Eqs.\ (\ref{tau}), (\ref{f})
\begin{equation}
\tau(1)=0.74\pm 0.15~,~~~~~\rho^2=0.90\pm 0.05~.
\label{linear2}
\end{equation}
The errors come from the QCD sum rule parameters, the continuum threshold and
the Borel parameter.

\section{Results and Discussions}

The decay rate for $B\to PT$ is in general given by
\begin{equation}
\Gamma(B\to PT)=\frac{|{\vec p}_P|^5}{12\pi m_T^2}\Bigg(\frac{m_B}{m_T}\Bigg)^2
\Bigg|\frac{{\cal A}(B\to PT)}{\epsilon^*_{\mu\nu}p_B^\mu p_B^\nu}\Bigg|^2~,
\end{equation}
where ${\vec p}_P$ is the pseudoscalar three momentum.
We use the WCs of \cite{Oh,Dutta}, where 
$c_1(m_b)=1.149$ and $c_2(m_b)=-0.3209$.
The branching ratios for $B\to D_2^*\pi$ are summarized in Table \ref{t1}
with the abbreviations
\begin{mathletters}
\begin{eqnarray}
\calB_{+-}&\equiv&\calB(\Bbar^0\to D_2^{*+}\pi^-)~,\\
\calB_{0-}&\equiv&\calB(B^-\to D_2^{*0}\pi^-)~.
\end{eqnarray}
\end{mathletters}
In Table \ref{t1}, we give the numerical results for various $\xi$, 
while it is reported that $N_C^{\rm eff}(B\to D\pi)\approx 2$ \cite{HYC}.
As a comparison, results from the ISGW2 model \cite{jplee} are also given.
The QCD sum rule predicts about four-times larger branching ratios.
Since the decay amplitude is proportional to $a_1$, the dependence of $\calB$
on $\xi$ is rather mild.
\par
\begin{figure}
\vskip 2cm
\begin{center}
\epsfig{file=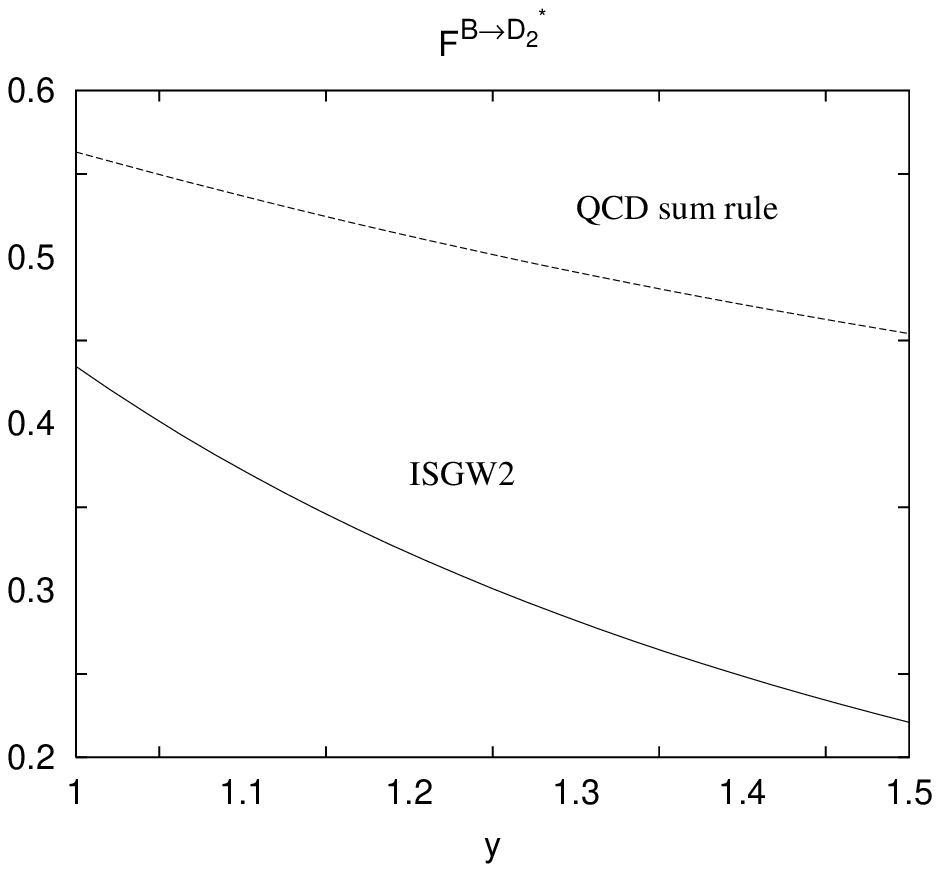}
\end{center}
\caption{}
\label{F0}
\end{figure}

Figure \ref{F0} shows the transition form factors as functions of $y$ for
both QCD sum rule and the ISGW2, where the form factors $F^{B\to T}$ are 
defined by
\begin{mathletters}
\begin{eqnarray}
{\cal T}&=&i\frac{G_F}{\sqrt{2}}V_{cb}V^*_{ud}f_P\epsilon^*_{\mu\nu}
p_B^\mu p_B^\nu F^{B\to T}(m_P^2) a_1~,\\
F^{B\to T}(m_P^2)&=&k(m_P^2)+(m_B^2-m_T^2)b_+(m_P^2)+m_P^2b_-(m_P^2)~.
\end{eqnarray}
\end{mathletters}
\par
With the linear approximation of (\ref{linear}) and (\ref{linear2}), 
the ratio of the hadronic and semileptonic branching ratios is (with only 
central values of (\ref{linear2}))
\begin{equation}
\frac{\calB(B\to D_2^*\pi)}{\calB(B\to D_2^*\ell{\bar\nu})}=
0.23~(\xi=0.1)~,~0.20~(\xi=0.3)~,0.18~(\xi=0.5)~,
\label{nlsl}
\end{equation}
where the estimated total decay rate 
$\Gamma(B\to D_2^*\ell{\bar\nu})=2.11\times 10^{-15}~{\rm GeV}$ \cite{Huang} 
is used.
In this fraction, common factor of $|\tau(1)|^2$ is canceled and only the slope
parameter $\rho^2$ remains at leading order.
Near future experiments will check the ratio.
One thing interesting is that the value is
very close to that of $B\to D_1$ transition \cite{Neubert}
\begin{equation}
\frac{\calB(B^-\to D_1^0\pi^-)}{\calB(B^-\to D_1^0\ell^-{\bar\nu})}=0.21\pm0.08.
\end{equation}
\par
Another important ratio is 
\begin{eqnarray}
R_{21}^{\rm had}&\equiv& \frac{\calB(B\to D_2^*\pi)}{\calB(B\to D_1\pi)}
\nonumber\\
&=&
\left(\frac{r_1}{r_2^3}\right)
\left(\frac{p_2^5}{m_B^2 p_1^3}\right)
\left[\frac{\tau(y_2)}{\tau(y_1)}\right]^2
\left[\frac{2r_2(1+y_2)-(1-r_2^2)+r_\pi^2}
     {(1-y_1^2)-3(1-r_1y_1)+(y_1-2)(y_1-r_1)}\right]^2~,
\label{R21}
\end{eqnarray}
where
\begin{mathletters}
\begin{eqnarray}
p_{1,2}&\equiv& \frac{1}{2m_B}\sqrt{[m_B^2-(m_{D_1,D_2^*}+m_\pi)^2]
        [m_B^2-(m_{D_1,D_2^*}-m_\pi)^2]}~,\\
y_{1,2}&\equiv& \frac{m_B^2+m_{D_1,D_2^*}^2-m_\pi^2}{2m_B m_{D_1,D_2^*}}~,
\end{eqnarray}
\end{mathletters}
and $r_{1,2,\pi}\equiv m_{D_1,D_2^*,\pi}/m_B$.
Charged decay modes of $B\to D_1(D_2^*)\pi$ are both highly dominated by tree 
amplitudes, and involve the same IW function at the heavy quark limit within 
the factorization scheme.
We have, at the heavy quark limit, 
\begin{equation}
R_{21}^{\rm had}=0.21\pm0.01~.
\end{equation}
This value is rather small compared to the semileptonic decay ratio 
\footnote{This leading order (at $m_Q\to\infty$) prediction is larger than 
experimental data.  See \cite{Huang}.}
\cite{Huang}
\begin{equation}
R_{21}^{\rm sl}\equiv 
\frac{\calB(B\to D_2^*\ell\bar{\nu})}{\calB(B\to D_1\ell\bar{\nu})}
=1.55~,
\end{equation}
and experiments 
$\calB(B^-\to D_2^{*0}\pi^-)/\calB(B^-\to D_1^0\pi^-)=1.8\pm 1.0$ (CLEO
\cite{Neubert}), $0.89\pm0.14$ (Belle \cite{Belle2}).
Since the mass difference between $D_1$ and $D_2^*$ is very small 
($m_{D_1}/m_{D_2^*}\approx 0.99$), (\ref{R21}) is quite insensitive to the
structure of the leading IW function; $\tau(y_2)/\tau(y_1)\approx 1$.
The ratio $R_{21}^{\rm had}$ depends on how the relevant form factors are
described in HQET, so the subleading analysis might affect the ratio.
The authors of \cite{Leibovich} found that $R_{21}^{\rm had}$ depends on 
$\tau$ weakly but is very sensitive to the subleading IW functions, ranging
from $0$ to $1.5$ while Neubert in \cite{Neubert} expects $1/m_Q$ corrections
less than 20\%, with the leading order prediction 
$R_{21}^{\rm had}\approx 0.35$.
While the experimental errors are still large (much reduced in Belle),
the discrepancy between theory and data is significant.
More quantitative and reliable analysis at subleading order in theory and
reduced errors in the future experiments will check this HQET framework 
as well as the factorization.
\par
\begin{figure}
\vskip 2cm
\begin{center}
\epsfig{file=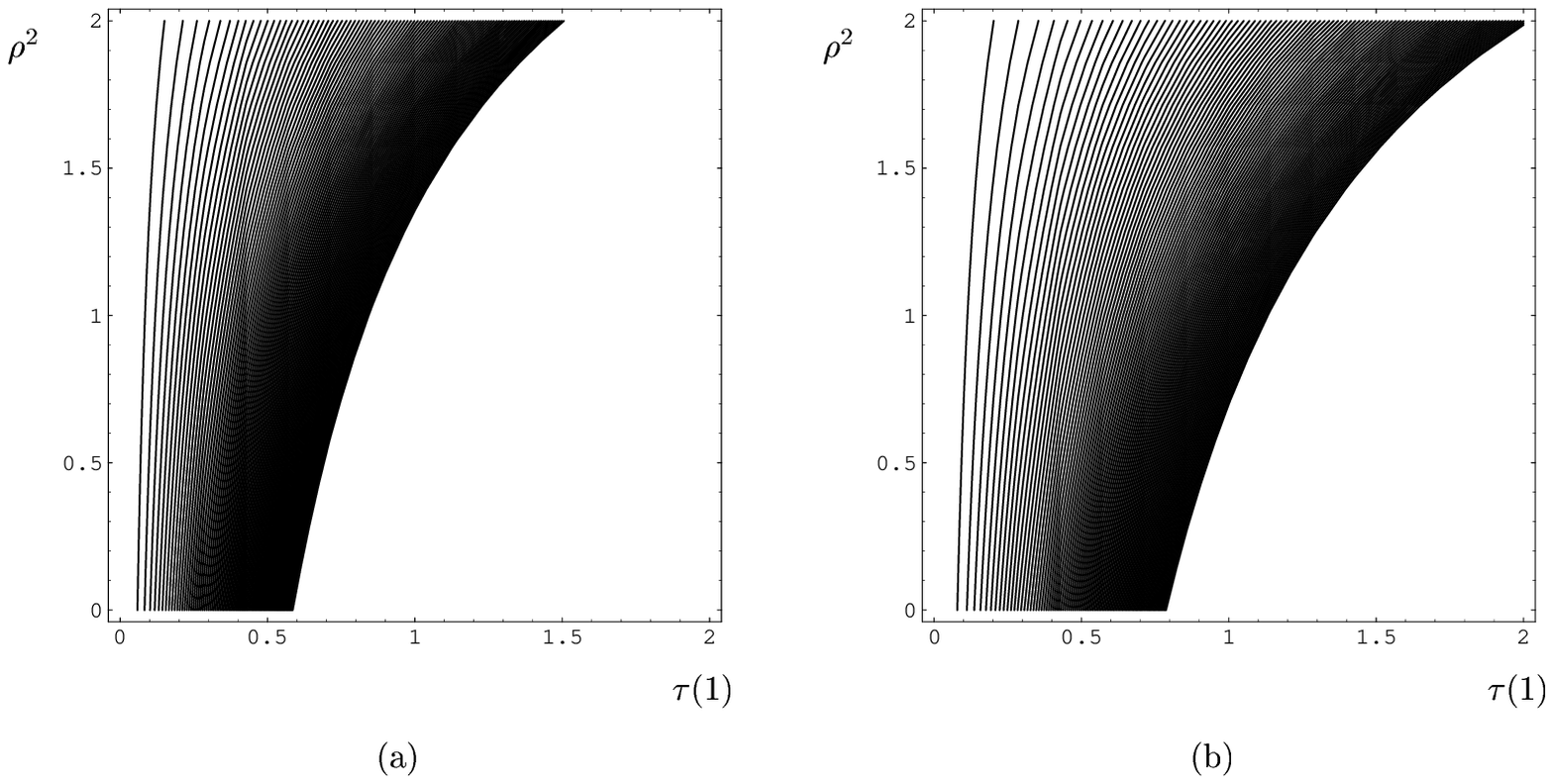}
\end{center}
\caption{}
\label{contour}
\end{figure}

On the other hand, conversely,
the experimental bounds on the branching ratios can constrain $\tau(1)$ and
$\rho^2$.
Figure \ref{contour} shows the allowed region (shaded) of these two parameters
from the given branching ratios.
Note that the central value of QCD sum rule result 
$(\tau(1),\rho^2)=(0.74,0.90)$ resides very near to the boundary.
In addition, measured value of the ratio (\ref{nlsl}) will determine $\rho^2$
regardless of $\tau(1)$.
More precise measurements will thus provide a simple test of leading order
description of $B\to D_2^*$.
\par
Now consider the next-to-leading order (NLO) corrections to the above analysis.
The NLO contributions come from both $\Lambda_{\rm QCD}/m_Q$ and $\alpha_s$.
In HQET, $\Lambda_{\rm QCD}/m_Q$ corrections appear in a two-fold way.
At the Lagrangian level, subleading terms are summarized in $\lambda_1$ and 
$\lambda_2$.
In a usual convention, $\lambda_1$ parametrizes the kinetic term of higher
order derivative, while $\lambda_2$ represents the chromomagnetic interaction
which explicitly breaks the heavy quark spin symmetry.
Their effects are known to be small \cite{Leibovich}.
At the current level, $\Lambda_{\rm QCD}/m_Q$ corrections come from the current
matching procedure onto the effective theory.
The correction terms originate from the small
portion of the heavy quark fields which corresponds to the virtual motion of
the heavy quark.
During the current matching, eight subleading IW functions are newly introduced
and two of them are independent.
\par
Compared with the $B\to D_1$ transition, $\Lambda_{\rm QCD}/m_Q$ corrections
to $B\to D_2^*$ would be rather small.
This is due to the tensor structure in the final state.
At zero recoil (properties at this point is very important because 
kinematically allowed region is quite narrow around it), 
the transition matrix is proportional to
\begin{equation}
\langle D_2^*(v,\epsilon)|j^\mu|B(v)\rangle\sim
\epsilon^{\mu\nu}v_\nu=0~.
\end{equation}
As argued briefly in the Introduction, vanishing matrix element is well
explained by the HQS at the heavy quark limit.
On the other hand, for $B\to D_1$ transition at zero recoil,
\begin{equation}
\langle D_1(v,\epsilon)|j^\mu|B(v)\rangle\sim\epsilon^{*\mu}~,
\end{equation}
which can be nonzero in general.
That's the reason why NLO corrections are more important in $B\to D_1$.
\par
Another NLO contribution from $\calO(\alpha_s)$ corrections is studied in
$B\to D'_0, D'_1$ decays in \cite{Colangelo}; 
it remains a good challenge in $B\to D_2^*$ process.
\par
Finally, if there observed a sizable value of 
$\calB_{00}\equiv\calB(\Bbar^0\to D_2^{*0}\pi^0)$,
it should come from the nonfactorizable effects.
Or indirectly, the measurement of $(\kappa\calB_{+-}-\calB_{0-})$ where
$\kappa\equiv\tau_{B^+}/\tau_{\Bbar^0}\approx 1.07$ is the $B$ life-time ratio,
would test the validity of the isospin relation (\ref{triangle}) and the general
factorization scheme.
The values in Table \ref{t1}, of course, satisfy the relation 
$\calB_{0-}/\calB_{+-}=\kappa$.

\section{Summary}

Using the QCD sum rule results for the leading IW function of $B\to D_2^*$,
we investigated the nonleponic two-body decays $B\to D_2^*\pi$ within the
framework of factorization.
The predicted branching ratios are about four times larger than the recent
calculations based on the ISGW2.
Study of tensor meson is very advantageous because the decay amplitudes are
simple, and some of the decay modes are directly related to the nonfactorizable
effects.
Present $B$ factories show a big optimism of producing copious tensor mesons,
and more precise and reliable theoretical works are on the request.
The NLO analysis of both $\calO(\Lambda_{\rm QCD}/m_Q)$ and 
$\calO(\alpha_s)$, in this respect, is challenging and will improve the
theoretical reliability.
\\[5mm]
\begin{center}
{\large\bf Acknowledgements}\\[5mm]
\end{center}

This work was supported by the BK21 Program of the Korea Ministry of Education.


\newpage

\begin{center}{\large\bf FIGURE CAPTIONS}\end{center}

\noindent
Fig.~1
\\
Form factors $F^{B\to D_2^*}(y)$ from QCD sum rule and ISGW2.
\vskip .3cm
\par

\noindent
Fig.~2
\\
Allowed regions (shaded) of $(\tau(1),\rho^2)$ from the experimental bounds
for (a) $B^-\to D_2^{*0}\pi^-$ and (b) $\Bbar^0\to D_2^{*+}\pi^-$.
\vskip .3cm
\par

\pagebreak

\pagebreak


\begin{table}
\caption{Branching ratios for $B\to D_2^*\pi$ with QCD sum rule 
(ISGW2 [6]).
The QCD sum rule results are from the linear approximation of Eq.\ 
(\ref{linear}).}
\begin{tabular}{c|ccc}
& $\xi=0.1$ & $\xi=0.3$ & $\xi=0.5$ \\\hline
$\calB_{+-}\times 10^4$ &11.42 (3.11)&10.14 (2.76)&8.94 (2.44) \\
$\calB_{0-}\times 10^4$ &12.17 (3.31)&10.81 (2.94)&9.53 (2.59) 
\end{tabular}
\label{t1}
\end{table}


\begin{thebibliography}{99}

\bibitem{Katoch}
A.C.\ Katoch and R.C.\ Verma, \prd{52}{1995}{1717}.

\bibitem{Munoz}
J,H.\ Mu\~noz. A.A.\ Fojas, and G.\ L\'opez Castro, \prd{59}{1999}{077504};
\ibid{55}{1997}{5581}.

\bibitem{Oh}
C.S.\ Kim, B.H.\ Lim, and Sechul Oh, \epjc{22}{2002}{683}; 
{\it ibid}. 695 (2002).

\bibitem{ISGW}
N.\ Isgur, D.\ Scora, B.\ Grinstein, and M.B.\ Wise, \prd{39}{1989}{799}.

\bibitem{Belle}
A.\ Garmash (Belle Collaboration), to appear in the Proc. of the 4th 
Internatioal Conference on $B$ Physics \& CP Violation, Ise-Shima, Japan, 
19-23 Feb, 2001 [hep-ex/0104018];
K.\ Abe \etal~ (Belle Collaboration), contributed to Lepton Photon 01, Rome, 
Italy, 23-28 Jul, 2001 [hep-ex/0107051];
A.\ Garmash \etal~ (Belle Collaboration), [hep/ex/0201007].

\bibitem{jplee}
C. S. Kim, Jong-Phil Lee, and Sechul Oh, [hep-ph/0205262];
[hep-ph/0205263].

\bibitem{ISGW2}
D.\ Scora and N.\ Isgur, \prd{52}{1995}{2783}.

\bibitem{Abe}
K.\ Abe \etal~ (Belle Collaboration), [hep-ex/0107048];
D.\ Cassel, talk given at the 20th International Symposium on Lepton and
Photon Interactions at High Energies (Lepton Photon 01), July 2001, Rome, Italy.

\bibitem{Xing}
Z.\ Xing, [hep-ph/0107257];
H.-Y.\ Cheng, \prd{65}{2002}{094012};
M.\ Neubert and A.A.\ Petrov, \plb{519}{2001}{50};
J.-P.\ Lee, [hep-ph/0109101].

\bibitem{Neubert}
M.\ Neubert, \plb{418}{1998}{173}.

\bibitem{Diehl}
M.\ Diehl and G.\ Hiller, \jhep{0106}{2001}{067}.

\bibitem{PDG}
Particle Data Group, D.E.\ Groom \etal, \epjc{15}{2000}{1}.

\bibitem{Leibovich}
A.K.\ Leibovich, Z.\ Ligeti, I.W.\ Stewart, and M.B.\ Wise,
\prl{78}{1997}{3995};
\prd{57}{1998}{308}.

\bibitem{Huang}
M.-Q.\ Huang and Y.-B.\ Dai, \prd{59}{1999}{034018}.

\bibitem{sumrule}
M.\ Shifman, A.\ Vainshtein and V.\ Zakharov, \npb{147}{1979}{385};
{\it ibid.}, 448 (1979).

\bibitem{Dutta}
B.\ Dutta and Sechul Oh, \prd{63}{2001}{054016}.

\bibitem{HYC}
H.-Y.\ Cheng and K.-C\ Yang, \prd{59}{1999}{092004}.

\bibitem{Belle2}
K.\ Abe \etal~ (Belle Collaboration), BELLE-CONF-0235.

\bibitem{Colangelo}
P.\ Colangelo, F.\ De Fazio, and N.\ Paver, \prd{58}{1998}{116005}.

\end{thebibliography}
\end{document}